\documentclass[9pt,twocolumn,twoside]{opticajnl}
\journal{opticajournal} % use for journal or Optica Open submissions

\setboolean{shortarticle}{true}
\usepackage{lineno}
\title{Tunable broadband polarization retarders}

\author[1]{Hristina S. Hristova}
\author[2]{Svetoslav S. Ivanov}
\author[2]{Nikolay V. Vitanov}
\author[2,*]{Andon A. Rangelov}

\affil[1]{Institute of Solid State Physics, Bulgarian Academy of Sciences, 72 Tsarigradsko chauss\'{e}e, 1784 Sofia, Bulgaria}
\affil[2]{Center for Quantum Technologies, Department of Physics, Sofia University, James Bourchier 5 Blvd, 1164 Sofia, Bulgaria}

\affil[*]{rangelov@phys.uni-sofia.bg}

\begin{abstract}
We theoretically propose a type of tunable polarization retarder, which is composed of sequences of half-wave and quarter-wave polarization retarders, allowing operation at broad spectral bandwidth. The constituent retarders are composed of stacked standard half-wave retarders and quarter-wave retarders rotated at designated angles relative to their fast-polarization axes.
The proposed composite retarder can be tuned to an arbitrary value of the retardance by varying the middle retarder alone, while maintaining its broadband spectral bandwidth intact.
\end{abstract}

\begin{document}

\maketitle

%%%%%%%%%%%%%%%%%%%%%%%%%%%%%%%%%%%%%%%%%%%%%%%%%%%%%%%%%%%%%%%%%%%%%%%%%%%%%%%%%%%%%%%%%%%%%%%%%%%%%%%%%%%%%%%%%%%%%%%%%%%%%%%

\section{Introduction}

%%%%%%%%%%%%%%%%%%%%%%%%%%%%%%%%%%%%%%%%%%%%%%%%%%%%%%%%%%%%%%%%%%%%%%%%%%%%%%%%%%%%%%%%%%%%%%%%%%%%%%%%%%%%%%%%%%%%%%%%%%%%%%%

One of light's fundamental properties is its polarization \cite{Hecht,Wolf,Azzam,Goldstein,Duarte}. 
The capacity to observe and control the polarization state proves highly advantageous for practical applications. Various optical measurement techniques grounded in polarization find application in stress measurements, ellipsometry, physics, chemistry, biology, astronomy, and other fields  \cite{Pye,Damask,Landolfi}. 
Additionally, the controlled manipulation of light polarization forms the foundational principle for display and telecommunication technologies \cite{Matioli}.

A key optical element for polarization state manipulation is the retarder, also
known as a wave plate. It introduces a controlled phase difference or
retardation between the orthogonal components of polarized light as it
passes through. The primary purpose of a retarder is to modify the
polarization state of light, and it is often employed to convert linearly
polarized light into elliptically or circularly polarized light, or vice
versa \cite{Hecht,Wolf,Azzam,Goldstein,Duarte}.

Retarders are categorized based on the specific way they alter the
polarization state and the amount of phase difference they introduce. The
most common types of retarders include quarter-wave plates and half-wave
plates.
A quarter-wave plate introduces a quarter-wavelength phase shift between the two orthogonal components of polarized light. It is particularly useful for converting linearly polarized light into circularly polarized light or vice versa.
A half-wave plate introduces a half-wavelength phase shift. It is often used to rotate the plane of polarization of linearly
polarized light.

Although half-wave and quarter-wave plates are widely used in practical applications, retarders with adjustable retardation values can also be specifically designed and utilized effectively. Given specific scenarios, some may require a half-wave plate while others a quarter-wave plate. Accordingly, the ability to adjust the retardance is crucial. Liquid-crystal
technology provides a means to achieve tunable retardance \cite{Sharp,Sit,Ye}. As an alternative, a method akin to Evans arbitrary retarder \cite{Evans49} was suggested by Messaadi \textit{et al.} \cite{Messaadi}, involving the cascading of two half-wave plates between two quarter-wave plates. This fundamental optical system
operates as an adjustable retarder, allowing control over the retardance by
rotating the two half-wave plates.

Both half-wave and quarter-wave plates show dispersion, i.e., the phase
shift (retardation) depends on the wavelength of the incident light.
However, many applications require achromatic (broadband) polarization retarders \cite%
{Ardavan,Ivanov,Peters}.

In this paper, we introduce a class of spectrally broadband half-wave and quarter-wave polarization retarders. This
conceptual framework involves combining conventional wave plates with the optical axis of individual retarders rotated at specific angles. Leveraging
the mathematical analogy between the description of polarization rotation
and the dynamics of a light-driven two-level quantum system (akin to the
concept of composite pulses \cite{Levitt}), we identified optimal rotation
sequences for the wave plates. In contrast to single wave plates, these
composite sequences can mitigate the chromatic dependence of
retardation, enabling operation at broad bandwidths. 

We then propose an \emph{arbitrarily} adjustable
retarder consisting of a composite half-wave plate sandwiched
between two parallel composite quarter-wave plates. By simply
rotating the composite half-wave plate, the retardance is continuously
adjusted. If the used composite wave plates are broadband then broadband arbitrary polarization retarder is created.  

%***************************************************************
\begin{figure}[thb]
\centerline{\includegraphics[width=0.95\columnwidth]{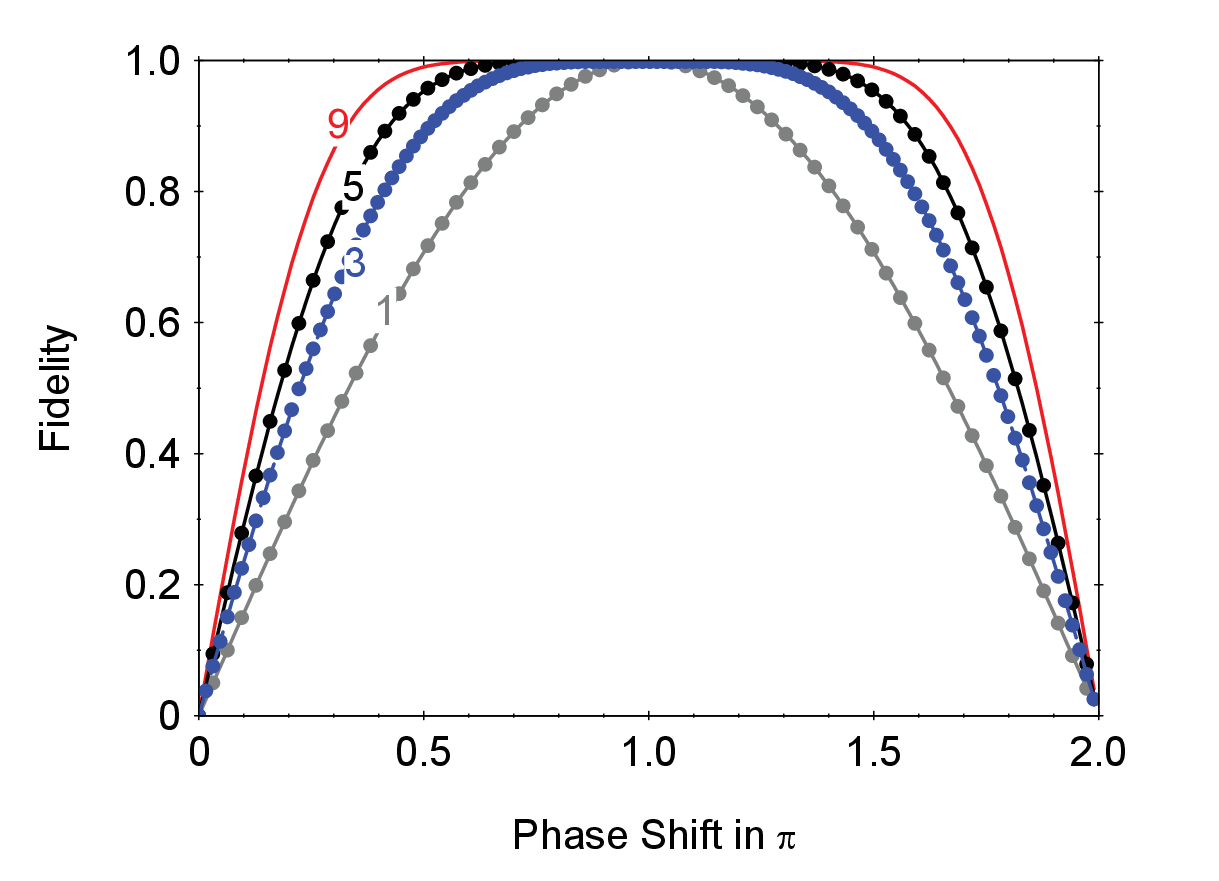}}
\caption{(Color online) Fidelity $F$ vs phase shift $\protect\varphi $ for
broadband half-wave retarders for different numbers of constituent plates $N$%
. The rotation angles are given in Table \protect\ref{Table}. The fidelities
of the single half-wave plate retarder is shown with labels \textquotedblleft
1\textquotedblright , and the composite cases are for \textquotedblleft
3\textquotedblright , \textquotedblleft 5\textquotedblright\ and
\textquotedblleft 9\textquotedblright\ wave plates series. As
one can see, longer retarder sequences offer a higher level of
stability against variations in the phase shift $\protect\varphi $. }
\label{Fig1}
\end{figure}
%***************************************************************
%***************************************************************
\begin{table}[hbt]
\begin{tabular}{|c|l|}
\hline
$N$ & Rotation angles $(\theta _{1}$; $\theta _{2}$; $\cdots $; $\theta
_{N-1}$; $\theta _{N}$) \\ \hline
& (a) broadband half-wave retarders \\ \hline
3 & (129; 65; 115) \\ 
5 & (90; 142; 67; 67; 142) \\ 
9 & (0; 158; 60; 45; 152; 174; 108; 0; 59) \\ \hline\hline
& (b) broadband quarter-wave retarders \\ \hline
4 & (90; 159; 49; 159) \\
5 & (0; 229; 146; 146; 229) \\
6 & (90; 27; 159; 265; 159; 27) \\ \hline\hline
\end{tabular}%
\caption{Rotation angles $\protect\theta _{k}$ (in degrees) for composite broadband retarders with varying numbers $N$ of constituent
wave plates. (a) broadband half-wave composite retarders with Jones matrix $\mathbf{J}^{(N)}=\mathbf{J}_{\protect\theta _{N}}(\protect\pi )\mathbf{J}_{\protect\theta _{N-1}}(\protect\pi )\cdots \mathbf{J}_{\protect\theta _{1}}(\protect\pi )$; (b) broadband quarter-wave composite  retarders with Jones
matrix $\mathbf{J}^{(N)}=\mathbf{J}_{\protect\theta _{N}}(\protect\pi )\mathbf{J}_{\protect\theta _{N-1}}(\protect\pi )\cdots \mathbf{J}_{\protect\theta _{1}}(\protect\pi /2)$. In accordance with the usual experimental inaccuracy, all rotation angles are rounded to the nearest degree (1 degree accuracy).}
\label{Table}
\end{table}
%***************************************************************

%%%%%%%%%%%%%%%%%%%%%%%%%%%%%%%%%%%%%%%%%%%%%%%%%%%%%%%%%%%%%%%%%%%%%%%%%%%%%%%%%%%%%%%%%%%%%%%%%%%%%%%%%%%%%%%%%%%%%%%%%%%%%%%

\section{Tunable arbitrary retarder}

%%%%%%%%%%%%%%%%%%%%%%%%%%%%%%%%%%%%%%%%%%%%%%%%%%%%%%%%%%%%%%%%%%%%%%%%%%%%%%%%%%%%%%%%%%%%%%%%%%%%%%%%%%%%%%%%%%%%%%%%%%%%%%%

A rotator with a rotation angle $\theta $ is described by the Jones
matrix
\begin{equation}
\mathbf{R}(\theta )=\left[
\begin{array}{cc}
\cos \theta & \sin \theta \\
-\sin \theta & \cos \theta%
\end{array}%
\right] ,
\end{equation}%
while the Jones matrix for a retarder is
\begin{equation}
\mathbf{J}(\varphi )=\left[
\begin{array}{cc}
e^{i\varphi /2} & 0 \\
0 & e^{-i\varphi /2}%
\end{array}%
\right] .
\end{equation}%
Here $\varphi $ is the phase shift between the two orthogonal polarizations. 
%These unitary operations are known in quantum computation as rotation and phase quantum gates \cite{Nielsen2000}.
The quarter-wave plate ($\varphi =\pi /2$) and the
half-wave plate ($\varphi =\pi $) are the most often used retarders \cite%
{Pye,Damask}. In a frame, where the horizontal and the vertical
polarization axes are rotated at an angle $\theta $ with respect to the slow
and fast axes of the plate, the retarder is described by
\begin{equation}
\mathbf{J}_{\theta }(\varphi )=\mathbf{R}(-\theta )\mathbf{J}(\varphi )%
\mathbf{R}(\theta ).  \label{retarder}
\end{equation}

The following discussion concentrates on the construction of a configurable retarder, as proposed by Messaadi \textit{et al.} \cite{Messaadi}. 
The idea is that a rotator positioned between two crossed quarter-wave plates functions as an arbitrary
retarder, with the retardation defined by the rotator's rotation angle \cite{Ye}.
The rotator can be obtained by a set of two half-wave plates \cite%
{Zhan,Rangelov} with a relative angle of $\alpha /2$ between their fast polarization axes (up to an unimportant minus sign):
\begin{equation}
\mathbf{J}_{\theta }(\pi )\mathbf{J}_{\theta +\alpha /2}(\pi )=-\left[
\begin{array}{cc}
\cos \alpha & \sin \alpha \\
-\sin \alpha & \cos \alpha%
\end{array}%
\right] .  \label{rotator}
\end{equation}%s
The configurable retarder thus takes the form \cite{Messaadi}
\begin{equation}
\mathbf{J}_{0}(\alpha )=-\mathbf{J}_{-\frac{\pi }{4}}(\pi /2)\mathbf{J}%
_{\theta _{1}}(\pi )\mathbf{J}_{\theta _{1}+\frac{\alpha }{4}}(\pi )\mathbf{J%
}_{\frac{\pi }{4}}(\pi /2).
\end{equation}%
We can simplify this into Evans's retarder \cite{Evans49} by dividing a half-wave plate into two quarter-wave plates,
\begin{equation}
\mathbf{J}_{\theta _{1}}(\pi )=\mathbf{J}_{\theta _{1}}(\pi /2)\mathbf{J}%
_{\theta _{1}}(\pi /2).
\end{equation}%
Setting up the free parameter $\theta _{1}=\pi /4$ we get
\begin{equation}
\mathbf{J}_{0}(\alpha )=-\mathbf{J}_{-\frac{\pi }{4}}(\pi /2)\mathbf{J}_{\frac{\pi }{4}%
}(\pi /2)\mathbf{J}_{\frac{\pi }{4}}(\pi /2)\mathbf{J}_{\frac{\pi }{4}+\frac{%
\alpha }{4}}(\pi )\mathbf{J}_{\frac{\pi }{4}}(\pi /2).  \notag
\end{equation}%
After taking into account that
%\begin{equation}
$\mathbf{J}_{-\frac{\pi }{4}}(\pi /2)\mathbf{J}_{\frac{\pi }{4}}(\pi /2)=\hat{%
1},$
%\end{equation}%
we find that, up to an unimportant sign, we can construct a configurable retarder using two quarter-wave plates and a half-wave plate, rotated at a particular angle:
\begin{equation}
\mathbf{J}_{0}(\alpha )=-\mathbf{J}_{\frac{\pi }{4}}(\pi /2)\mathbf{J}_{\frac{\pi }{4}+%
\frac{\alpha }{4}}(\pi )\mathbf{J}_{\frac{\pi }{4}}(\pi /2).  \label{arbitrary-to-arbitrary polarization}
\end{equation}%
%This is analogous to the construction of a phase gate in quantum computation by the Hadamard gates sandwiching a rotation gate \cite{Nielsen2000}.

Several remarks about this device are noteworthy. First, it features fewer optical
elements compared to Messaadi et al. \cite{Messaadi}. 
Therefore, if will feature less reflection and absorption losses.
Second, it is very convenient to use since the retardation is produced by just
rotating the half-wave plate. Additionally, this system is universal in that
any waveplate, even broadband waveplates (like Fresnel rhombs) may be used
to create it; in this instance, the suggested device would also be
broadband. 

The theory for composite half- and quarter-wave plates operating in the
broadband regimes will be developed in the next section and the result will be applied in Eq. (\ref{arbitrary-to-arbitrary
polarization}) to obtain tunable broadband retarders.

%%%%%%%%%%%%%%%%%%%%%%%%%%%%%%%%%%%%%%%%%%%%%%%%%%%%%%%%%%%%%%%%%%%%%%%%%%%%%%%%%%%%%%%%%%%%%%%%%%%%%%%%%%%%%%%%%%%%%%%%%%%%%%%

\section{Composite wave plates}

%%%%%%%%%%%%%%%%%%%%%%%%%%%%%%%%%%%%%%%%%%%%%%%%%%%%%%%%%%%%%%%%%%%%%%%%%%%%%%%%%%%%%%%%%%%%%%%%%%%%%%%%%%%%%%%%%%%%%%%%%%%%%%%

Building retarders that are insensitive to phase shift $\varphi
$ at a chosen value of this shift is our first goal. In order to do this, we
replace the single retarder with a series of $N$ retarders, each of which
has a phase shift of $\varphi _{k}$ and is rotated by an angle of $\theta
_{k}$, as represented by the total Jones matrix (read from right to left).
\begin{equation}
\mathbf{J}^{\left( N\right) }=\mathbf{J}_{\theta _{N}}\left( \varphi
_{N}\right) \mathbf{J}_{\theta _{N-1}}\left( \varphi _{N-1}\right) \cdots
\mathbf{J}_{\theta _{1}}\left( \varphi _{1}\right) .
\label{overall Jones matrix}
\end{equation}%
We use Ardavan's definition of fidelity of a matrix from polarization optics \cite{Ardavan} to compute the efficiency of the composite retarder:
\begin{equation}
F=\frac{1}{2}\text{Tr}\left( \mathbf{J}_{0}^{-1}\mathbf{J}^{(N)}\right) ,
\label{fidelity}
\end{equation}%
where $\mathbf{J}^{(N)}$ and $\mathbf{J}_{0}$ represent the achieved and the target retarders, respectively. If the two matrices are
identical then $F=1$, but the fidelity reduces if they differ.

%%%%%%%%%%%%%%%%%%%%%%%%%%%%%%%%%%%%%%%%%%%%%%%%%%%%%%%%%%%%%%%%%%%%%%%%%%%%%%%%%%%%%%%%%%%%%%%%%%%%%%%%%%%%%%%%%%%%%%%%%%%%%%%

\subsection{Broadband half-wave retarders}

%%%%%%%%%%%%%%%%%%%%%%%%%%%%%%%%%%%%%%%%%%%%%%%%%%%%%%%%%%%%%%%%%%%%%%%%%%%%%%%%%%%%%%%%%%%%%%%%%%%%%%%%%%%%%%%%%%%%%%%%%%%%%%%

We now demonstrate how to build a \emph{broadband half-wave retarder} that
can accurately invert light polarization across a wide range of phase shifts
$\varphi $ about $\pi $. It is achieved with an odd number of half-wave plates ($\varphi
_{k}=\varphi =\pi $), each rotated by an angle $\theta _{k}$, where the angles are
considered free parameters.
The Jones matrix for the sequence is
\begin{equation}
\mathbf{J}^{(N)}=\mathbf{J}_{\theta _{N}}(\pi )\mathbf{J}_{\theta
_{N-1}}(\pi )\cdots \mathbf{J}_{\theta _{1}}(\pi )
\end{equation}%
and we set $\mathbf{J}^{(N)}=\left[
\begin{array}{cc}
i & 0 \\
0 & -i%
\end{array}%
\right]$ at $\varphi =\pi $, this
means that $N-1$ independent angles $\theta _{k}$ remain to be fixed. Then,
for every $\varphi =\pi $, we nullify as many lowest order derivatives of $%
\mathbf{J}_{11}^{(N)}$ versus the phase shift $\varphi $ as we can. As a
result, for the $N-1$ rotation angles $\theta _{k}$, we derive a set of
nonlinear algebraic equations. The first $\lfloor (N-1)/2\rfloor $ complex
derivatives, where $\lfloor x\rfloor $ denotes the integer part of $x$, may
be nullified by $N-1$ rotation angles:
\begin{equation}
\left[ \partial _{\varphi }^{k}\mathbf{J}_{11}^{\left( N\right) }\right]
_{\varphi =\pi }=0\quad \left( k=1,2,...,\lfloor (N-1)/2\rfloor \right) .
\label{nullify11}
\end{equation}%
Solutions to \eqref{nullify11} provide broadband half-wave retarders.
Greater order of stability against fluctuations in the phase shift $\varphi $
and the light wavelength $\lambda $ is provided by longer retarder
sequences, with a greater number $N$ of constituent wave plates. Table \ref%
{Table} contains examples of broadband half-wave retarders and their
fidelities are shown in Fig.~\ref{Fig1}.

%***************************************************************
\begin{figure}[htb]
\centerline{\includegraphics[width=0.95\columnwidth]{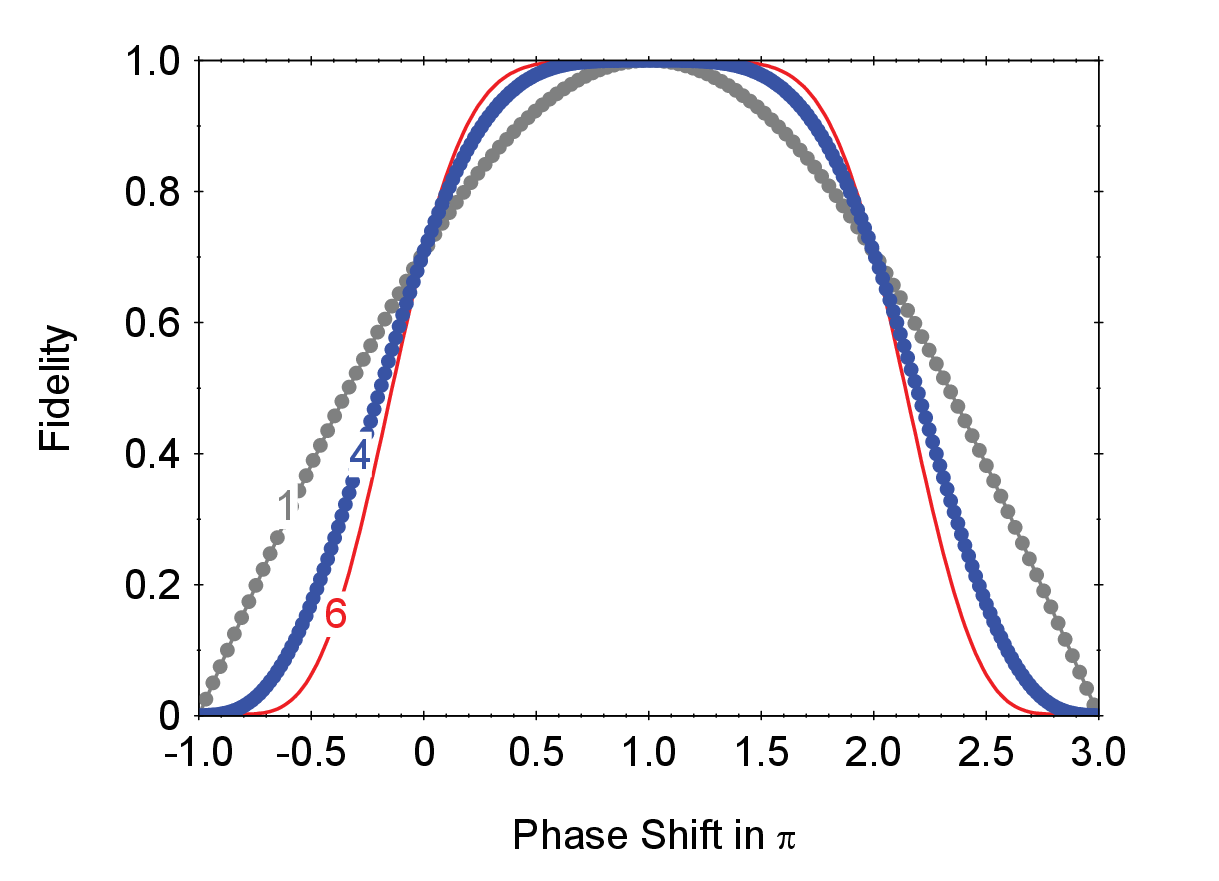}}
\caption{(Color online) Fidelity $F$ vs phase shift $\protect\varphi $ for
broadband quarter-wave retarders, for different numbers of constituent
plates $N$. The rotation angles are given in Table \protect\ref{Table}. The
fidelities of the single-plate retarder is shown with labels
\textquotedblleft 1\textquotedblright\ and the composite cases are for
\textquotedblleft 4\textquotedblright and \textquotedblleft 6\textquotedblright . As one can
clearly see, longer retarder sequences offer a higher level of stability
against variations in the phase shift $\protect\varphi $. %\textcolor{red}{These are passband curves.} 
}
\label{Fig3}
\end{figure}
%***************************************************************

%***************************************************************
\begin{figure}[htb]
\centerline{\includegraphics[width=1\columnwidth]{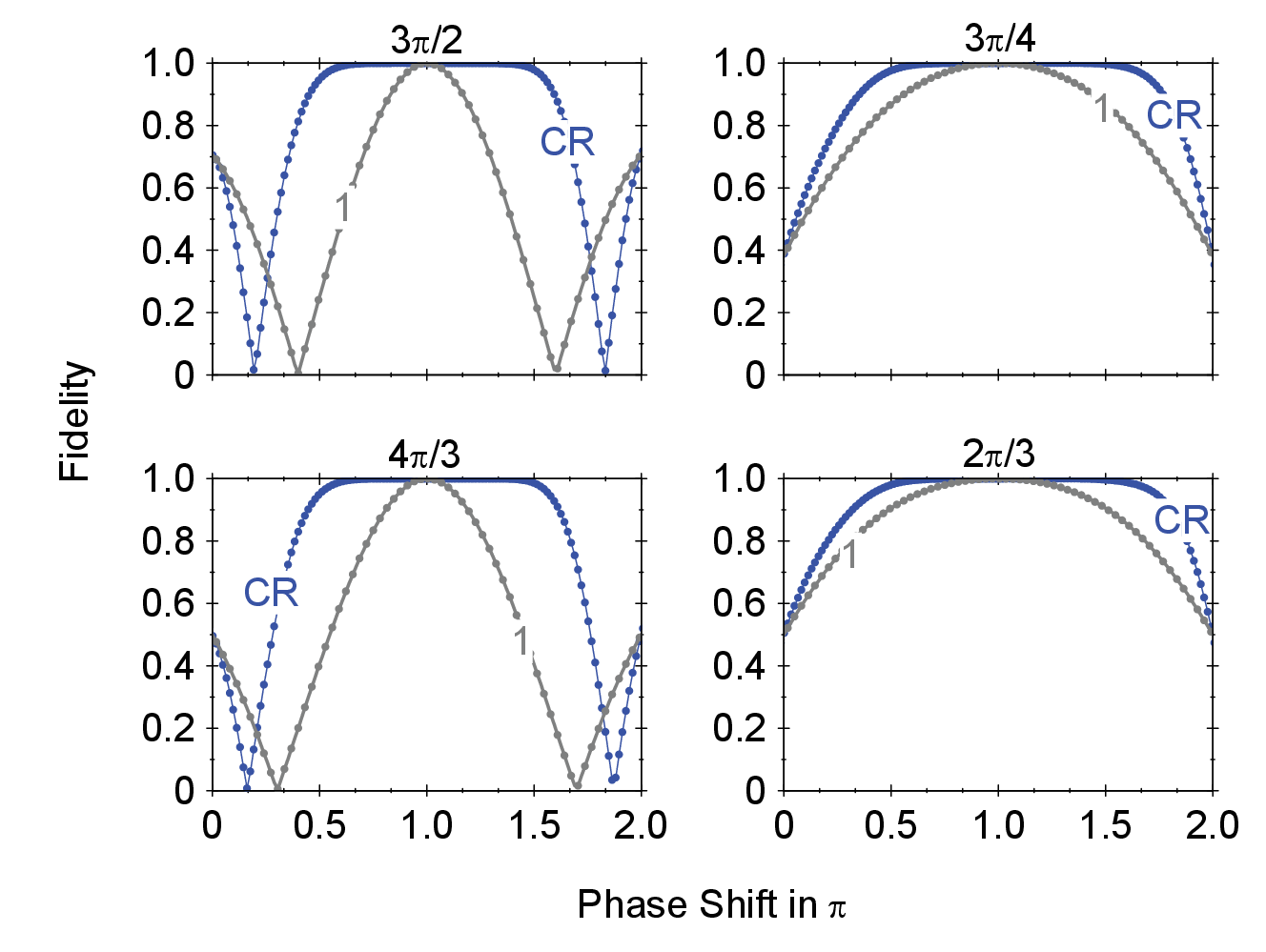}}
\caption{(Color online) Fidelity $F$ vs phase shift $\protect\varphi $ for
composite broadband arbitrary retarder, for different retardations $\frac{3}{%
2}\protect\pi ,\frac{4}{3}\protect\pi ,\frac{3}{4}\protect\pi $ and $\frac{2%
}{3}\protect\pi $. The performance of arbitrary retarder composed by
combining two standard quarter-wave plates and one standard half-wave plate
is shown with labels \textquotedblleft 1\textquotedblright\ where the
arbitrary composite retarder is labeled CR. }
\label{Fig5}
\end{figure}
%***************************************************************

%%%%%%%%%%%%%%%%%%%%%%%%%%%%%%%%%%%%%%%%%%%%%%%%%%%%%%%%%%%%%%%%%%%%%%%%%%%%%%%%%%%%%%%%%%%%%%%%%%%%%%%%%%%%%%%%%%%%%%%%%%%%%%%

\subsection{Broadband quarter-wave retarder}

%%%%%%%%%%%%%%%%%%%%%%%%%%%%%%%%%%%%%%%%%%%%%%%%%%%%%%%%%%%%%%%%%%%%%%%%%%%%%%%%%%%%%%%%%%%%%%%%%%%%%%%%%%%%%%%%%%%%%%%%%%%%%%%

\emph{Broadband quarter-wave retarders} can be built in a similar way to
\emph{broadband half-wave retarders}. Their Jones matrix of quarter-wave
plate is
\begin{equation}
\mathbf{J}_{0}(\pi /2)=\left[
\begin{array}{cc}
e^{i\pi /4} & 0 \\
0 & e^{-i\pi /4}%
\end{array}%
\right].
\end{equation}%
We have found that in this case, the optimal composite sequence consists of
one quarter-wave plate ($\varphi =\pi /2$) and $N-1$ half-wave plates ($%
\varphi =\pi $). The Jones matrix that corresponds to this is
\begin{equation}
\mathbf{J}^{(N)}=\mathbf{J}_{\theta _{N}}(\pi )\mathbf{J}_{\theta
_{N-1}}(\pi )\cdots \mathbf{J}_{\theta _{2}}(\pi )\mathbf{J}_{0}(\pi /2).
\end{equation}%
Using the $N-1$ free phases $\theta _{k}$, we can cancel the first $\lfloor
(N-1)/2\rfloor $ complex derivatives.
\begin{equation}
\left[ \partial _{\varphi }^{k}J_{11}^{\left( N\right) }\right] _{\varphi
=\pi }=0\quad (k=1,2,\ldots ,\lfloor (N-1)/2\rfloor ).  \label{nullify21}
\end{equation}%
We can additionally eliminate the real or imaginary part of the subsequent
non-zero derivative (of order $N/2$) for even $N$:
\begin{equation}
\text{Re}\left[ \partial _{\varphi }^{N/2}\mathbf{J}_{11}^{\left( N\right) }%
\right] _{\varphi =\pi }=0\text{ or }\text{Im}\left[ \partial _{\varphi
}^{N/2}\mathbf{J}_{11}^{\left( N\right) }\right] _{\varphi =\pi }=0.
\label{nullify22}
\end{equation}%
Table \ref{Table} provides a list of phases that result in quarter-wave
retarders with broadband capabilities and their fidelities are shown in Fig.~%
\ref{Fig3}.

%%%%%%%%%%%%%%%%%%%%%%%%%%%%%%%%%%%%%%%%%%%%%%%%%%%%%%%%%%%%%%%%%%%%%%%%%%%%%%%%%%%%%%%%%%%%%%%%%%%%%%%%%%%%%%%%%%%%%%%%%%%%%%%

\section{Broadband tunable arbitrary retarder}

%%%%%%%%%%%%%%%%%%%%%%%%%%%%%%%%%%%%%%%%%%%%%%%%%%%%%%%%%%%%%%%%%%%%%%%%%%%%%%%%%%%%%%%%%%%%%%%%%%%%%%%%%%%%%%%%%%%%%%%%%%%%%%%

So far we have shown how to build broadband half- and
quarter-wave plates. Now combining composite half-wave plate sandwiched
between two parallel composite quarter-wave plates using Eq. (\ref%
{arbitrary-to-arbitrary polarization}), we achieve broadband arbitrary retarder as it was hinted earlier.

First, let's build a broadband arbitrary polarization retarder using series
of six wave plates to construct each broadband composite quarter-wave plates and
series of nine wave plates to build the broadband composite half-wave plate,
where each series is given in Table \ref{Table}. The performance of such
composite retarder (CR) is given in Fig.~\ref{Fig5} for several retardation
cases ( $\frac{3}{2}\pi ,\frac{4}{3}\pi ,\frac{3}{4}\pi $ and $\frac{2}{3}%
\pi $). A scrutiny of Fig. \ref{Fig5} clearly shows that composite retarder
performance is much more broadband compared to the case where we have used
standard wave plates in Eq. (\ref{arbitrary-to-arbitrary polarization}). 
The fidelity profiles might potentially be made arbitrarily flat by increasing the number of retarders in the composite series.
However, the practical application of such many-retarders sequences is uncertain because of the multiple optical components in the composite series; using commercial achromatic wave plates would be a more viable addition in this scenario. 
%\textcolor{red}{NV: Can we have this device with fewer plates?}

%%%%%%%%%%%%%%%%%%%%%%%%%%%%%%%%%%%%%%%%%%%%%%%%%%%%%%%%%%%%%%%%%%%%%%%%%%%%%%%%%%%%%%%%%%%%%%%%%%%%%%%%%%%%%%%%%%%%%%%%%%%%%%%

\section{Summary}

%%%%%%%%%%%%%%%%%%%%%%%%%%%%%%%%%%%%%%%%%%%%%%%%%%%%%%%%%%%%%%%%%%%%%%%%%%%%%%%%%%%%%%%%%%%%%%%%%%%%%%%%%%%%%%%%%%%%%%%%%%%%%%%

In conclusion, one can achieve an arbitrarily adjustable retarder
consisting of a half-wave plate sandwiched between two parallel quarter-wave
plates. By simply rotating the half-wave plate, the retardance is
continuously adjusted. The idea is universal since arbitrary broadband retarders may be achieved using achromatic wave plates. Our method of creating broadband half and
quarter-wave retarders, involves utilizing the nuclear magnetic resonance analog of composite pulses. Since commercial
wave plates may be employed, an experimental implementation of the proposed device should be possible in most optical laboratories.

%%%%%%%%%%%%%%%%%%%%%%%%%%%%%%%%%%%%%%%%%%%%%%%%%%%%%%%%%%%%%%%%%%%%%%%%%%%%%%%%%%%%%%%%%%%%%%%%%%%%%%%%%%%%%%%%%%%%%%%%%%%%%%%%%%%%%%%%%%%%%%%%%%%%
\section{Acknowledgments}
This research is partially supported by the Bulgarian national plan for recovery and resilience, contract BG-RRP-2.004-0008-C01 SUMMIT: Sofia
University Marking Momentum for Innovation and Technological Transfer, project number 3.1.4. 

\section{Disclosures} The authors declare no conflicts of interest.

\end{document}